\PassOptionsToPackage{dvipsnames}{xcolor}
\documentclass[fleqn,10pt]{wlscirep}
\usepackage[english]{babel}
\usepackage[utf8]{inputenc}
\usepackage[T1]{fontenc}
\usepackage{doi}
\usepackage{textcomp}
\usepackage{float}
\usepackage{siunitx}
\usepackage{import}
\usepackage{mathdots}
\usepackage{nicefrac}
\usepackage[ruled]{algorithm2e}
\usepackage{setspace}
\usepackage{placeins}
\usepackage{multicol}
\usepackage{multirow}
\usepackage{array}
\newcolumntype{P}[1]{>{\centering\arraybackslash}p{#1}}
\usepackage{tabularx}
\usepackage{longtable}
\usepackage{pgfplots}
\pgfplotsset{compat=newest}
\usepgfplotslibrary{external}
\usetikzlibrary{external}
\usetikzlibrary{shapes.geometric, shapes.arrows}
\usetikzlibrary{decorations.pathmorphing, decorations.pathreplacing, calligraphy, decorations.shapes, decorations.markings}
\usetikzlibrary{calc}
\usetikzlibrary{fit}
\usetikzlibrary{patterns}
\usetikzlibrary{positioning}
\usetikzlibrary{arrows, arrows.meta}
\tikzset{
    position/.style args={#1:#2 from #3}{
        at= (#3.#1), anchor=#1+180, shift= (#1:#2)
    }
}
\newif\ifcomptikz
\comptikzfalse

\usepackage{acro}
\DeclareAcroEnding{possessive}{'s}{'s}
\NewAcroCommand\acg{m}{\acropossessive\UseAcroTemplate{first}{#1}}
\NewAcroCommand\acsg{m}{\acropossessive\UseAcroTemplate{short}{#1}}
\NewAcroCommand\aclg{m}{\acropossessive\UseAcroTemplate{long}{#1}}

\DeclareAcronym{NDT}{
short = NDT,
long = nondestructive testing
}

\DeclareAcronym{PSF}{
short = PSF,
long = point spread function
}

\DeclareAcronym{PPT}{
short = PPT,
long = pulsed phase thermography
}

\DeclareAcronym{FMC}{
short = FMC,
long = full matrix capture
}

\DeclareAcronym{FWHM}{
short = FWHM,
long = full width at half maximum
}

\DeclareAcronym{ROI}{
short = ROI,
long = region of interest,
long-plural-form = regions of interest
}

\DeclareAcronym{FFT}{
short = FFT,
long = fast Fourier transform
}

\DeclareAcronym{PSR}{
short = photothermal SR,
long =  photothermal super resolution
}

\DeclareAcronym{SR}{
short = SR,
long = super resolution
}

\DeclareAcronym{SNR}{
short = SNR,
long = signal-to-noise ratio
}

\DeclareAcronym{ADMM}{
short = ADMM,
long = alternating direction method of multipliers
}

\DeclareAcronym{OUT}{
short = OuT,
long = object under test
}

\DeclareAcronym{NMSE}{
short = NMSE,
long = normalized mean square error
}

\DeclareAcronym{SSIM}{
short = SSIM,
long = structural similarity
}

\DeclareAcronym{DLP}{
short = DLP,
long = digital light processing
}

\DeclareAcronym{DMD}{
short = DMD,
long = digital micromirror device
}

\DeclareAcronym{MWIR}{
short = MWIR,
long = midwave infrared
}

\newcommand{\appropto}{\mathrel{\vcenter{
  \offinterlineskip\halign{\hfil$##$\cr
    \propto\cr\noalign{\kern2pt}\sim\cr\noalign{\kern-2pt}}}}}

\newcommand{\reffig}[1]{Figure~\ref{#1}}
\newcommand{\reftab}[1]{Table~\ref{#1}}
\newcommand{\refeqq}[1]{Equation~(\ref{#1})}
\newcommand{\refalgol}[1]{Algorithm~(\ref{#1})}

\newcommand{\di}{defect/inhomogeneity}
\newcommand{\dip}{defects/inhomogeneities}

\DeclareMathOperator*{\minimize}{minimize}
\DeclareMathOperator*{\argmin}{arg\,min}
\DeclareMathOperator*{\argwhere}{arg\,where}
\DeclareMathOperator*{\NMSE}{NMSE}

\title{Nondestructive thermographic detection of internal defects using pixel-pattern based laser excitation and photothermal super resolution reconstruction}

\author[1,*]{Julien Lecompagnon}
\author[1]{Philipp Daniel Hirsch}
\author[2]{Christian Rupprecht}
\author[1]{Mathias Ziegler}
\affil[1]{Bundesanstalt für Materialforschung und -prüfung (BAM), Thermographic Methods, Berlin, 12205, Germany}
\affil[2]{Technische Universität Berlin, Chair of Coating Technology, Berlin, 10623, Germany}
\affil[*]{\href{mailto:julien.lecompagnon@bam.de}{julien.lecompagnon@bam.de}}


\begin{abstract}
    In this work, we present a novel approach to photothermal super resolution based thermographic resolution of internal defects using two-dimensional pixel pattern-based active photothermal laser heating in conjunction with subsequent numerical reconstruction to achieve a high-resolution reconstruction of internal defect structures. With the proposed adoption of pixelated patterns generated using laser coupled high-power DLP projector technology the complexity for achieving true two-dimensional super resolution can be dramatically reduced taking a crucial step forward towards widespread practical viability. Furthermore, based on the latest developments in high-power DLP projectors, we present their first application for structured pulsed thermographic inspection of macroscopic metal samples. In addition, a forward solution to the underlying inverse problem is proposed along with an appropriate heuristic to find the regularization parameters necessary for the numerical inversion in a laboratory setting. This allows the generation of synthetic measurement data, opening the door for the application of machine learning based methods for future improvements towards full automation of the method. Finally, the proposed method is experimentally validated and shown to outperform several established conventional thermographic testing techniques while conservatively improving the required measurement times by a factor of 8 compared to currently available photothermal super resolution techniques.
\end{abstract}

\begin{document}

\flushbottom
\maketitle

\acuse{PSR}
\section*{Introduction}\label{sec:intro}
Active thermographic testing as a \ac{NDT} method is a very efficient technique for contactlessly detecting surface defects as well as \dip{} well below the surface of any \ac{OUT}. In active thermographic testing the \ac{OUT} is actively extrinsically heated, its temperature evolution is recorded and the resulting change in temperature is then evaluated to detect irregularities, which imply possible \dip{}. While the capabilities of thermographic testing is constrained by the utilized hardware like the spatial and temperature resolution of the infrared camera used, thermographic testing is also fundamentally constrained by the diffusive nature of heat propagation unlike other methods, which rely on propagating waves such as ultrasonic testing or radiography. When it comes to the detection/resolution of internal \dip{} deep below the surface, empirically it has been shown, that the ratio between the depth at which a \di{} occurs and its spatial extension should be close to unity for it to be fully resolved~\cite{Burgholzer2022}. Exceeding this limit is one of the major challenges in modern thermographic testing and can be referred to as \ac{SR} thermography.

Even though thermographic \ac{SR} techniques are used in various fields within thermographic \ac{NDT}, e.g., for testing of biomaterials~\cite{Bouzin2019}, it is currently still mainly constrained to the spatial resolution enhancement of the utilized infrared cameras themselves~\cite{Raimundo2021,Sakagami2009}. Established defect resolution enhancing \ac{SR} methods are currently either only usable for the reconstruction for one-dimensional defects~\cite{Ahmadi2020,Ahmadi2021} or only approximate fully two-dimensional resolution enhancement by adding up multiple one-dimensionally structured illuminations~\cite{Burgholzer2017}. Recently, we have been able to expand the method to a true fully two-dimensional reconstruction utilizing sequential scanning with a single laser spot as the photothermal heat source~\cite{Lecompagnon2022}. While this technique already produces quite good results, the resulting measurement times are unfavorably high.

Within this work we show, how this issue can be overcome by the application of fully two-dimensionally structured illumination patterns using a laser-coupled \ac{DLP}-projector. The use of \ac{DLP} projectors in thermographic testing so far has been constrained to mostly inline shape recognition within different fields of application~\cite{An2016,Grubisic2011}. Only in a very limited form they have already been applied as an excitation source for thermal wave based detection~\cite{Pribe2016,Thiel2017}. Due to the recent advancements in output power for laser-coupled \ac{DLP}-based projectors mainly driven by the additive manufacturing industry, it is now feasibly possible for them to be applied as illumination sources for photothermal heating of \acp{OUT}.

\section*{Motivation: Photothermal Super Resolution Reconstruction}\label{sec:psr_maths}
The front surface temperature \(T_\text{meas}(x,y,z=0,t)\) of an \ac{OUT} experiencing a pulsed external photothermal heating with a pulse length \(t_\text{pulse}\) can be modelled in a Green's function like form as follows:
\begin{equation}\label{eq:T0+psf}
    T_\text{meas}(x,y,z=0,t) = T_0(x,y) + \Phi_\text{PSF}(x,y,t) *_{x,y}\, a(x,y)\ .
\end{equation}
Here, \(T_0(x,y)\) resembles the initial temperature of the \ac{OUT} at \(t=\SI{0}{\second}\)\,, \(\Phi_\text{PSF}(x,y,t)\) is the thermal \ac{PSF} characteristical for the \ac{OUT} and \(a(x,y)\) is the heat source distribution at play. The \ac{PSF} constitutes the response to a spatially Dirac-like heating of the \acg{OUT} front surface and can be defined analytically for the special case of a thermally thin plate knowing the material properties (thermal diffusivity \(\alpha\), density \(\rho\), specific heat capacity \(c_p\)) and geometry (plate thickness \(L\)) of the \ac{OUT} as follows~\cite{Cole2010}:
\begin{equation}\label{eq:psf}
    \Phi_{\text{PSF}}(x,y,t) = \left(\frac{2\,\mathrm{\hat{Q}}}{c_p\rho{(4\mathrm{\pi}\alpha t)}^{\nicefrac{n_\text{dim}}{2}}}\cdot \mathrm{e}^{-\frac{{(x-\bar{x})}^2+{(y-\bar{y})}^2}{4\alpha t}} \cdot \sum_{n=-\infty}^{\infty} R^{2n+1}\mathrm{e}^{-\frac{{(2nL)}^2}{4\alpha t}} \right) *_t\, I_t(t) .
\end{equation}
\(\mathrm{\hat{Q}}\) defines the amplitude of the external photothermally applied heat flux, \(n_\text{dim}\) the dimensionality of the heat flow (\(n_\text{dim}=3\) for a point-like heating), \(\left(\bar{x},\bar{y}\right)\) is the coordinate centroid, \(R\) the reflection coefficient for the thermal wave at the plate boundaries (for metals typically \(R\approx 1\)), \(I_t(t)\) the temporal structure of the external heating (typically a rectangular function of length \(t_\text{pulse}\)) and \(*_t\) indicates a convolution operation in time. The heat source distribution \(a(x,y)\) on the other hand consists of two parts, namely the external heat source distribution \(a_\text{ext}(x,y)\), which is a distribution of Dirac-pulses encoding the position at which the external heating is acting onto the \ac{OUT} and the internal heat source distribution part \(a_\text{int}\), which encapsulates the internal \di{} structure. Summing both parts and convolving the sum with the spatial structure of the external heating \(I_{x,y}(x,y)\) leads to the heat source distribution \(a(x,y)\) as follows:
\begin{equation}
    \label{eq:a_structure}
    a(x,y) = I_{x,y}(x,y) *_{x,y}\, \left(a_\text{ext}(x,y)+a_\text{int}(x,y)\right)\ ,
\end{equation}
where \(*_{x,y}\) denotes the convolution operation in both planar spatial dimensions. The internal heat source distribution in this context can be imagined as a distribution of \guillemotleft{}apparent\guillemotright{} heat sources, which can be phenomenologically described by the fact that a \di{} below the surface heated from above will impede the heat flow locally and therefore lead to a visible hot spot in the front surface temperature evolution, which qualitatively appears similar to how an active heat source embedded in the \ac{OUT} would~\cite{Lecompagnon2022}. This internal heat source distribution consists similarly to the external heat source distribution \(a_\text{ext}\) of a distribution of unit Dirac-pulses attenuated by the corresponding contrast factor \(\zeta\in\left[ 0,1\right[\) of the individual \di{} whose exact value depends on the effusivity contrast and depth of the defect:
\begin{equation}
    a_\text{int}(x,y)=\sum_i \zeta_i \cdot \delta(x_i,y_i)\ .
\end{equation}
The ultimate goal of \ac{PSR} reconstruction is then to solve \refeqq{eq:T0+psf} for the internal heat source distribution \(a_\text{int}\) and therefore acquiring a defect map of the examined \ac{ROI} on the \ac{OUT}. This reconstruction technique then achieves \ac{SR} capabilities by performing multiple measurements \(m\in \{1,\dots, n_m \}\) with varying external heating~\(a_\text{ext}\):
\begin{equation}
    \label{eq:core}
    \Phi_\text{PSF}(x,y,t) *_{x,y}\, a^m(x,y) =  T_\text{diff}^m(x,y,t)\ ,
\end{equation}
with \(T_\text{diff}^m(x,y,t)=T_\text{meas}^m(x,y,t)-T_0^m(x,y)\). This greatly increases the available information content about the internal defect structure since every measurement \(m\) contains the response of the \ac{OUT} and its \dip{} to a variety of different heating conditions and local heat flux directions.

However, in order to still be able to extract the effect of the internal \di{} structure independently of the external heating the following condition needs to be fulfilled:
\begin{equation}\label{eq:SRcond}
    I_{x,y}(x,y) *_{x,y}\, \sum_{m=1}^{n_m} a^m_\text{ext}(x,y) \approx const.
\end{equation}
This very important condition ensures, that on average every part of the \ac{ROI} is heated evenly and any deviations from the mean can be attributed to the heat flow impedance by the internal \dip{}. Since \refeqq{eq:T0+psf} is a severely ill-posed inverse problem, solving for \(a_\text{int}\) is not trivially possible. An approximative solution \(a_\text{rec}\) to this reconstruction problem can be determined by solving the following minimization problem that makes use of \(\ell_{2,1}\) and \(\ell_2\)-regularization, which incorporate prior information about the \di{} structure to restrict the solution space:
\begin{equation} \label{eq:min}
    \minimize_{a_\text{rec}}\, \frac{1}{2} \left\Vert\Phi_\text{PSF}(x,y,t) *_{x,y}\ a_\text{rec}^m(x,y) - T_\text{diff}^m(x,y,t)\right\Vert_2^2 + \lambda_{2,1}\left\Vert a_\text{rec}^m(x,y)\right\Vert_{2,1} + \lambda_2\left\Vert a_\text{rec}^m(x,y)\right\Vert_2^2\ ,
\end{equation}
where \(\left\Vert a_\text{rec}\right\Vert_{2,1}\) is the \(\ell_{2,1}\)-norm defined as:
\begin{equation}
    \left\Vert a_\text{rec}^m(x,y)\right\Vert_{2,1} = \sum_m\sqrt{\sum_{x,y} |{a_\text{rec}^m(x,y)}|^2} \ .
\end{equation}
The regularizer coefficients \(\lambda_{2,1}\) and \(\lambda_{2}\) govern the overall strength of the regularization and need to be inputted by the user. Currently, those factors have to be still determined empirically for each individual testing scenario since no definitive algorithm for automated determination is known. However, there exists recent ongoing work to solve this issue using machine learning techniques~\cite{Ahmadi2022,Hauffen2022}.

The severely ill-posed inversion problem in \refeqq{eq:min} can be solved numerically using the iterative \ac{ADMM} algorithm~\cite{Boyd2010} in the frequency domain as proposed in previous work~(for a detailed explanation of the inversion process see~[\citen{Lecompagnon2022}]).

\section*{Projection of two-dimensionally Structured Patterns}\label{sec:exp_trafo}
Most currently established experimental implementations of \ac{PSR} reconstruction are based on the sequential heating of the \ac{ROI} by projecting single spots or lines in a predefined grid pattern as the external photothermal heating~\cite{Burgholzer2017,Lecompagnon2022}. This kind of structured heating can be easily performed using only basic tooling but comes at the great disadvantage, that in order to cover a large \ac{ROI} a lot of independent measurements are necessary. Furthermore, there exists the possibility to make use of interference patterns (e.g.\ laser speckle patterns) as two-dimensionally structured illumination patterns, but those are mostly suited for materials/parts, which can be sufficiently heated with the rather low optical irradiances this technique provides~\cite{Burgholzer2017a}. For the simplified one-dimensional \ac{PSR} reconstruction technique, this problem has been already addressed by combining several laser lines as a heat source in order to cover a larger subsection of the \ac{ROI} per individual measurement~\cite{Ahmadi2021}. In this work we expand on this idea of combining several single excitations into a fully two-dimensionally structured illumination in order to achieve a significant improvement in the measurement times necessary.

The major disadvantage of using single laser spot excitation lies in the fact that each individual illumination only generates information about the \ac{OUT} in the near vicinity around the projected laser spot. If now multiple simultaneous laser spot excitations are combined into one single illumination, a significant reduction of the amount of illuminations necessary \(n_m\) can be accomplished.

\begin{figure}[!h]
    \centering
    \ifcomptikz
        \scalebox{1}{\import{figures/}{exp_trafo.tex}}
    \else
        \includegraphics[scale=1]{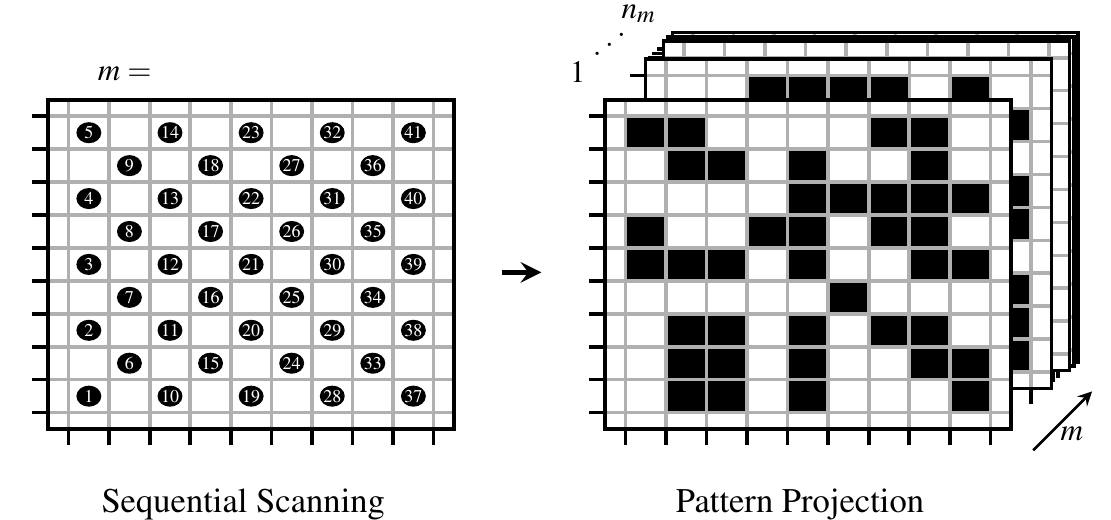}
    \fi
    \caption{Experimental Transformation: The left image shows qualitatively the arrangement of measurement positions (black dots) in sequential laser scanning based \ac{PSR} reconstruction. Here the \ac{ROI} is covered by \(n_m\) measurements where each measurement only covers a small subsection of the total \ac{ROI}. The right image shows the proposed new experimental technique of projecting a total of \(n_m\) different pixel patterns, which individually already span the whole \ac{ROI}. The black colored pixels are photothermally active (turned on).}\label{fig:exp_trafo}
\end{figure}

If the combined spots are arranged in an evenly spaced rectangular grid with a grid spacing similar to the spot diameter, then the resulting pattern can be thought of as a pixelized binary pattern, where every grid position (pixel) is either photothermally active (turned on) or not (turned off) as illustrated in \reffig{fig:exp_trafo}. Each of those patterns can then be further described by their pixel size \(d_\text{pix}\) and their fill factor \(\beta \in [0,1]\) where \(\beta=\nicefrac{n_\text{pix,on}}{n_\text{pix,total}}\), which is the ratio of photothermally active pixels \(n_\text{pix,on}\) and the total amount of pixels in the pattern \(n_\text{pixel}\).\\

\clearpage
While for sequential laser scanning the necessary amount of measurements for scanning a \ac{ROI} with area \(A_\text{ROI}\) is proportional to \(n_{m\text{,seq}}\appropto \nicefrac{A_\text{ROI}}{d_\text{spot}^2}\,\), in the limit, the projection of binary pixel patterns can reduce this down to the following requirement:
\begin{equation}\label{eq:n_m_min}
    n_{m\text{,pattern}} = \left\lceil\frac{1}{\beta}\right\rceil \ .
\end{equation}
This holds true as long as the pattern covers the whole \ac{ROI}. Even though a large \(\beta\) will lead to a lower number of measurements, it has to be still considered that for the \ac{PSR} reconstruction method to achieve super resolution capabilities, a three-dimensional heat flow is necessary within the \ac{OUT}. This is only the case for a sufficiently small \(\beta\) and \(d_\text{pix}\) compared to the expected defects to be resolved~\cite{Lecompagnon2022a}.\\

\section*{Illumination Pattern Design}\label{sec:pattern_design}
\begin{figure}[!h]
    \centering
    \ifcomptikz
        \scalebox{1}{\import{figures/}{pattern_overview.tex}}
    \else
        \includegraphics[scale=1]{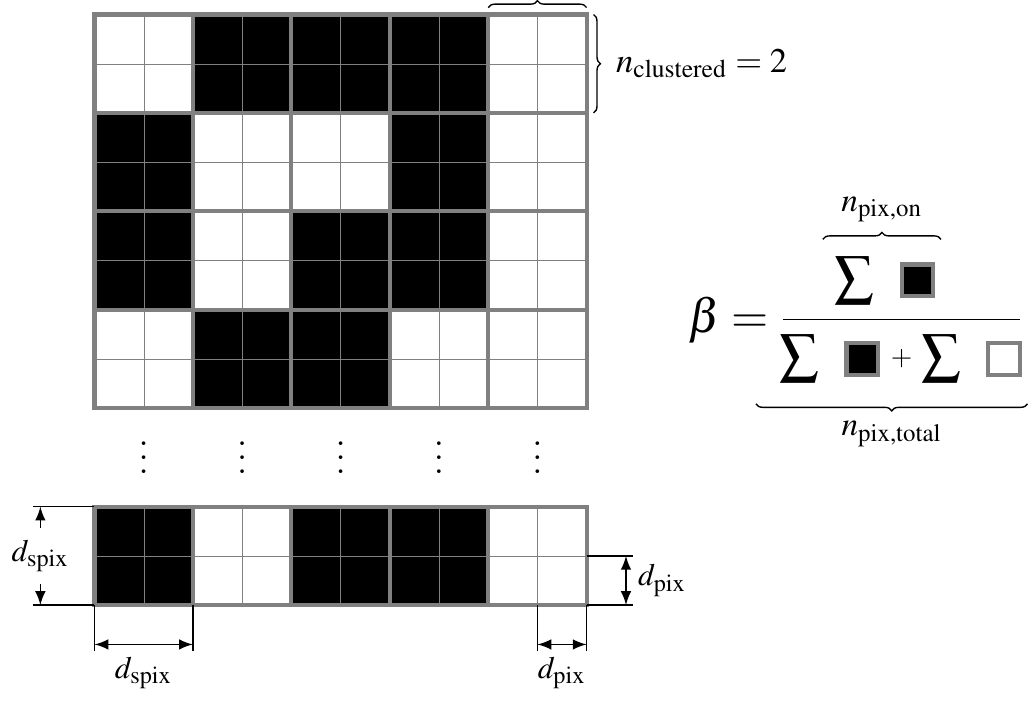}
    \fi
    \caption{Overview over the different parameters of the proposed random illumination patterns. Every pattern consists of \(n_\text{pix,total}\) individually addressable pixels with width \(d_\text{pix}\), which can be grouped together in larger \({n_\text{clustered}\times n_\text{clustered}}\) clusters with width \(d_\text{spix}\). The fill factor \(\beta\) can then be calculated as the ratio of the number of  activated pixels (\(n_\text{pix,on}\) shown in black color) and the total amount of pixels \(n_\text{pix,total}\) and is independent of the clustering.}\label{fig:pattern_overview}
\end{figure}

For the creation of the projected patterns, a random sampling approach is utilized within this work. This is a well known strategy in the field of compressed sensing and helps to find a suitable subset of patterns from the overall set of all possible patterns, which in general is by far too large to test all combinations of. Nevertheless, it is still necessary for the patterns used in \ac{PSR} reconstruction to fulfill the homogeneity constraint stated in \refeqq{eq:SRcond}. This constraint will be asymptotically fulfilled by random patterns for large \(n_{m} \gg \left\lceil\frac{1}{\beta}\right\rceil\). However, for smaller \(n_m\) the homogeneity constraint must be directly considered when constructing the patterns. Therefore, in this work we propose the adaption of a pseudo-random pattern generation strategy as presented in \refalgol{alg:pattern}, which is largely inspired by the famous Bresenham's algorithm in computer graphics~\cite{Bresenham1965}. Within this algorithm, all \(n_m\) patterns are created in sequence. For each newly created pattern only those pixels are taken into consideration for activation, which are currently activated less than expected for the given fill factor \(\beta\). If this subset of pixels is not large enough to reach the desired fill factor then the remaining pixels are activated at random until the desired fill factor is reached.
\begin{center}
    \begin{minipage}{0.9\linewidth}
        \begin{algorithm}[H]
            \begin{multicols}{2}
                \SetKwInOut{Input}{input}\SetKwInOut{Output}{output}
                \Input{\ \(\beta \in \mathbb{R}\), \(n_\text{pix,total}\), \(n_m \in \mathbb{N}\)}
                \Output{\ \(a_\text{ext} \in \{0,1\}^{n_\text{pix,total}\times\, n_m}\)}
                \begin{onehalfspacing}
                    \SetKwFunction{Ffill}{\(\text{fill}\)}
                    \SetKwProg{Fn}{function}{:}{}
                    \begin{singlespacing}
                        \nl \Fn{\Ffill{\(x \in \{0\}^{n_\text{pix,total}},\ n_\text{fill}\)}}{
                        \vspace{1mm}
                        \nl \(x \leftarrow x\) \parbox[t]{0.5\linewidth}{\(\text{filled uniformly at random with} \) \\ \( \text{at most}\ n_\text{fill}\ \text{ones}\)} \\
                        \nl \KwRet{} \(x\)}
                    \end{singlespacing}
                    \BlankLine{}
                    \nl \(n_\text{target} \leftarrow  \left\lceil\beta\cdot n_\text{pix,total}\right\rceil\)\\
                    \nl \(a^0_\text{ext} \leftarrow [0, \dots, 0]\) \\
                    \nl \(a^0_\text{ext} \leftarrow\) \texttt{fill}\(\left(a^0_\text{ext},\ n_\text{target}\right)\)\\
                    \BlankLine{}
                    \nl \For{\(m \leftarrow 2\) \KwTo{} \(n_m\)}{
                    \nl \(a^m_\text{ext} \leftarrow [0, \dots, 0]\) \\
                    \nl \(\text{share}_\text{on} \leftarrow \nicefrac{1}{m} \sum_{i=1}^m  a^i_\text{ext} \)\\
                    \nl\(\text{share}_\text{on,low} \leftarrow \argwhere\left(\text{share}_\text{on} < \beta\right)\)\\
                    \nl \(a^m_\text{ext}[\text{share}_\text{on,low}] \leftarrow \)\texttt{fill}\(\left(a^m_\text{ext}[\text{share}_\text{on,low}],\ n_\text{target}\right)\)\\
                    \vspace{1mm}
                    \nl \(n^m_{\text{pix,on}}\leftarrow \sum_{i=0}^{n_\text{pix,total}} a^m_\text{ext}[i]\)\\
                    \nl \If{\(n^m_{\text{pix,on}} < n_\text{target}\)}{
                    \vspace{1mm}
                    \nl \(n_\text{target,left} \leftarrow n_\text{target} - n^m_{\text{pix,on}}\)\\
                    \nl \(a^m_\text{ext}[\neg \text{share}_\text{on,low}] \leftarrow \)\texttt{fill}\parbox[t]{0.5\linewidth}{\(\left(a^m_\text{ext}[\neg \text{share}_\text{on,low}]\right.\), \\ \(\left.\hspace{1mm} n_\text{target,left}\right)\)}
                    }
                    }
                    \nl \KwRet{} \(a_\text{ext}\)\\
                \end{onehalfspacing}
            \end{multicols}
            \vspace*{0.7em}
            \caption{Pseudo-random pattern generation}\label{alg:pattern}
        \end{algorithm}
    \end{minipage}
\end{center}
Since combining an arbitrary arrangement of multiple laser spots is not very feasible in practice, a laser-coupled \ac{DMD}-based \ac{DLP} projector can be utilized instead. Those projectors feature the possibility to individually turn on and off any arbitrary single pixels in their output image, which has the same effect as combining a multitude of single laser spots. Modern \ac{DLP} projectors feature around \(10^6\) individual addressable pixels, which is more than enough for most use cases as a heat source for thermographic material testing. However, their biggest drawback is that currently commercially available \ac{DLP} projectors only feature at maximum a rather low optical output power of \({<\SI{100}{\watt}}\) for a fully-activated (\(\beta=1\)) image resulting in typical irradiances of \(5\text{-}\SI{25}{\watt\per\centi\metre\squared}\). Furthermore, this output power linearly decreases with the fill factor of the projected image, which additionally sets a lower bound for \(\beta\).\\
\clearpage
Due to the rather high pixel count of modern \ac{DLP} projectors, each pixel conveys only a tiny amount of the total optical output power. To deal with this issue, it is possible to group neighboring pixels into larger pixel clusters, which are then turned off and on in unison. For a grouping of \({n_\text{clustered}\times n_\text{clustered}}\) pixels, a new pixel cluster of side length \({d_\text{spix}=n_\text{clustered}\cdot d_\text{pix}}\) emerges while the total amount of available pixels is reduced to \(\nicefrac{n_\text{pix,total}}{n_\text{clustered}^2}\), which in turn increases the power per pixel to \({\hat{Q}_\text{total}\cdot \nicefrac{n_\text{clustered}^2\,}{n_\text{pix,total}}}\). To further illustrate the parameters of the pixel patterns and the grouping of pixels into clusters an overview is given in \reffig{fig:pattern_overview}.

\section*{Numerical Modeling: Forward Solution}\label{sec:fw_sol}
The underlying mathematical model behind the \ac{PSR} reconstruction approach is based on the inverse problem as stated in \refeqq{eq:T0+psf} for which an inversion for noisy measured data \(T_\text{meas}(x,y,t)\) in order to determine the internal heat source distribution \(a_\text{int}(x,y)\) (the defect map) is only possible using optimization algorithms. Due to the ill-posedness and the vastness of the solution space of the problem, any solution will always only be an approximation whose quality will be influenced by the proper choice of regularization as shown in \refeqq{eq:min}. However, for exploring the capabilities of the method numerically it would be highly beneficial to have the forward solution to the stated inverse problem. With the help of this forward solution it would then be possible to generate synthetic measurement data for a known \ac{OUT} with known internal defect distribution \(D(x,y)\). While such data can be obtained at high accuracy using finite-element simulations, this approach is also very computational expensive and it would be advantageous to have an approximative solution in closed form.

In order to find such an approximative forward solution, the heat source distribution \(a(x,y)\) and its parts \(a_\text{ext}(x,y)\) and \(a_\text{int}(x,y)\) have to be properly modelled. For the external heat source distribution \(a_\text{ext}(x,y)\), this already can be trivially achieved by inputting the external excitation pattern \(a_\text{pattern}\). For the internal heat source distribution \(a_\text{int}(x,y)\), a more sophisticated modeling is necessary. Due to the internal defects not being active heat sources as described in the phenomenological \guillemotleft{}apparent\guillemotright{} heat source explanation of the \ac{PSR} approach, the internal defect response is closely coupled to the external heating. This manifests itself in the fact, that the local strength of the internal heat source distribution \(a_\text{int}(x,y)\) is dependent on the relative positioning between the external excitation and the defect distribution \(D(x,y)\) in lateral (\(x,y\)) and also transversal (\(z\)) direction. The lateral positioning effect can be incorporated into the forward solution by element-wise multiplication of the defect distribution \(D(x,y)\) with the temperature field generated by the external heating \(\Phi_\text{PSF}(x,y) *_{x,y} a_\text{pattern}(x,y)\). This is necessary, since the internal defects can impede the heat flow as it is present at the defect location (no heating \(\rightarrow\) no signal).

The depth information of the defect can then be added to the model by introducing a numerical scaling factor \({\zeta \in \left[0,1\right[}\), which is attenuating the defect response according to the defect depth and effusivity contrast. Since this defect contrast factor is simplifying the involved physics of heat conduction quite substantially, it is quite hard to estimate and can be best determined by fitting the forward solution to empirical data generated by test measurements with sample defects at the desired depth.\\ % chktex 15

Overall this leads to the following equation, which can be used to generate synthetic measurement data \(T^m_\text{meas, sim}(x,y,t)\) for a given set of illumination patterns \(a^m_\text{pattern}\), a known defect distribution \(D(x,y)\), a known \ac{PSF} \(\Phi_\text{PSF}(x,y,t)\) and a suitable value for \(\zeta\) as follows:
\begin{equation}\label{eq:fws}
    \begin{split}
        T^m_\text{meas, sim}(x,y,t) &= \\
        \Phi_\text{PSF}(x,y,t)\, &*_{x,y}  \left( \underbrace{a_\text{pattern}^m(x,y)}_{I_{x,y}\, *_{x,y}\ a_\text{ext}} + \underbrace{\zeta \cdot  D(x,y) \odot \left( \Phi_\text{PSF}(x,y)  *_{x,y} a_\text{pattern}^m(x,y)\right)}_{a_\text{int}}                   \right) + T_0(x,y) + \mathcal{N}_\text{noise}(x,y) \ ,
    \end{split}
\end{equation}
where \(\odot\) denotes element-wise (Hadamard) multiplication and \(\mathcal{N}_\text{noise}(x,y)\) resembles an additional Gaussian measurement noise term.\\

\begin{figure}[!h]
    \centering
    \ifcomptikz
        \scalebox{1}{\import{figures/}{fit_qual_rsquared.tex}}
    \else
        \includegraphics[scale=1]{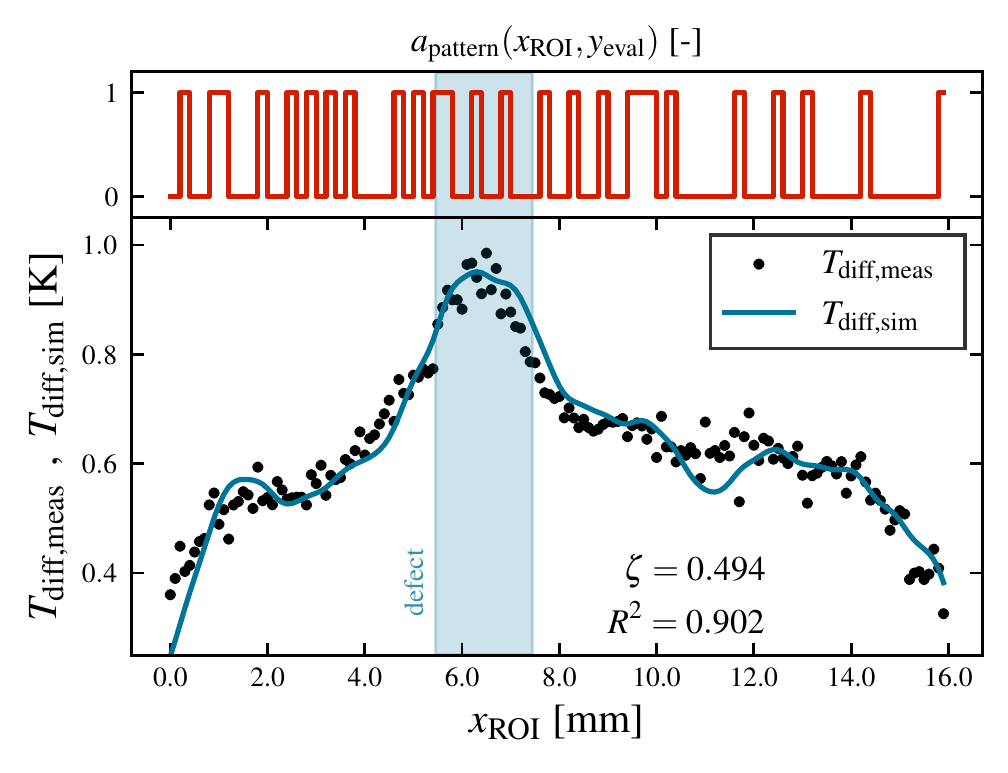}
    \fi
    \vspace{-1em}
    \caption{Quality of the forward solution and estimation of \(\zeta\): Synthetic measurement data \(T_\text{diff,sim}\) (blue line) generated for a \ac{ROI} featuring a \(\SI{2}{\milli\metre}\) wide defect \(\SI{0.5}{\milli\metre}\) below the surface (see \reffig{fig:sample_ROI}) and an illumination pattern with \({\beta=0.5}\) and \({d_\text{spix}=\SI{0.2}{\milli\metre}}\) is shown next to measured data \(T_\text{diff,meas}\) (black dots) using the same illumination pattern over the same \ac{ROI}. The defect contrast factor \({\zeta=0.494}\) has been determined as best-fit. The measured data and the synthetic data lie in good agreement (\({R^2=0.902}\)).}\label{fig:zeta_fit}
\end{figure}

For an exemplary test measurement on the \ac{OUT} shown in \reffig{fig:sample_ROI}, the performance of \refeqq{eq:fws} can be seen in \reffig{fig:zeta_fit}. Here, simulated measurement data for an illumination pattern with \({\beta=0.5}\) and \({d_\text{spix}=\SI{0.2}{\milli\metre}}\) is shown in comparison to measured data over the same \ac{ROI} using the same illumination pattern. The defect contrast is determined as best fit to \({\zeta=0.494}\) for the given defect size and depth within the \ac{ROI}. Even though the forward solution stated in \refeqq{eq:fws} only represents an approximation the model already features a high coefficient of determination of \({R^2>0.9}\) for wide ranges of tested measurement scenarios~\cite{Lecompagnon2021}.

\begin{figure}[!h]
    \centering
    \ifcomptikz
        \scalebox{1}{\import{figures/}{fit_qual_graph.tex}}
    \else
        \includegraphics[scale=1]{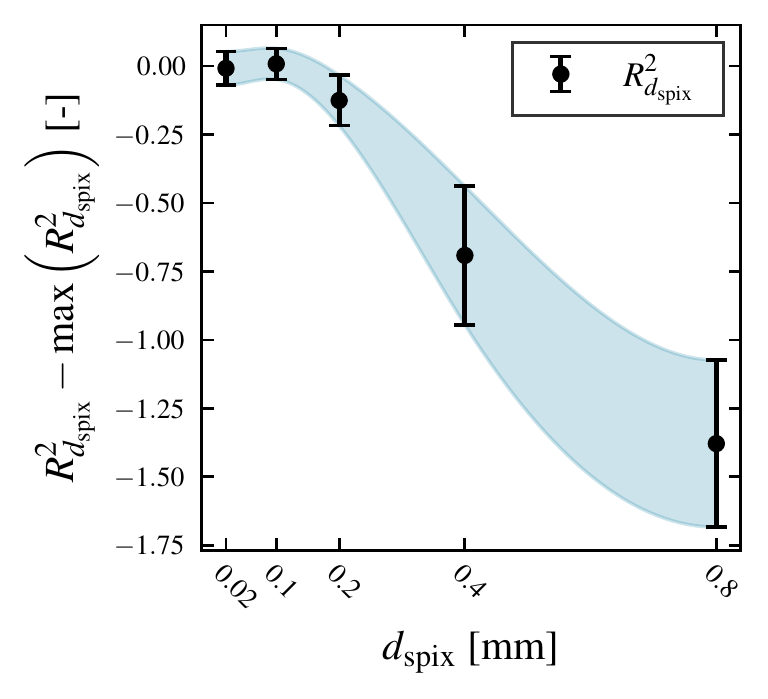}
    \fi
    \vspace{-1em}
    \caption{Sparsity constraint for 3D heat flow: Influence of the pixel cluster size \(d_\text{spix}\) on the quality of the presented forward model measured by the determination coefficient \(R_{d_\text{spix}}^2\). For each \(d_\text{spix}\) a total of \(n_m=20\) different patterns at \(\beta=0.5\) have been experimentally projected and the resulting temperature field has been compared to the prediction by \refeqq{eq:fws}. The presented data is normalized to the maximum achieved \({\max\left(R_{d_\text{spix}}^2\right)=0.723}\) value, which is calculated over the full \ac{ROI} including all edge effects and overlay errors. The shown error bars indicate \(\pm 1\sigma\) standard deviation.}\label{fig:sparsity_patterns}
\end{figure}

As already discussed in a previous section, the underlying model for \ac{PSR} reconstruction assumes a fully three-dimensional heat flow emerging in the \ac{OUT}. This sets the upper limit for the choice of \(\beta\) and \(d_\text{spix}\) (cf.~{[\citen{Almond1996}, p. 69]} setting the limit for a fully one-dimensional heat flow very conservatively at \(d_\text{spix} > 20 \cdot t_\text{diff}\)). For the choice of \(d_\text{spix}\) this dependency is investigated in \reffig{fig:sparsity_patterns}. Here, the coefficient of determination of the forward solution to a total of \(n_m=20\) different measurements for different illumination patterns with \({\beta=0.5}\) over the \ac{ROI} shown in \reffig{fig:sample_ROI} for five different values of \(d_\text{spix}\) is shown. In \reffig{fig:sparsity_patterns} it can be clearly seen, that for values above \({d_\text{spix}=\SI{0.2}{\milli\metre}}\) the fit quality is deteriorating quite fast and the deviation between different illumination patterns increases. A similar argument can be made for the fill factor \(\beta\).

\FloatBarrier
\section*{Reconstruction Quality and Automated Regularization Parameter Estimation}\label{sec:auto_param}
With the help of \refeqq{eq:fws} it is now possible to generate synthetic measurement data for numerical studies on the performance of \ac{PSR} reconstruction. For this to be feasible however, it is necessary to automate the generation and evaluation of different \ac{PSR} reconstruction results. Since the reconstructed internal heat source distribution \(a_\text{int}\) is sparse in nature as well as the defect distribution within realistic \acp{OUT}, all \ac{SNR}-based quality measures are not suited for evaluating the reconstruction quality. While there already exist quite sophisticated quality measures for determining the similarity between different distributions/images, which are also indiscriminant to small perturbations (e.g.\ the \ac{SSIM}~\cite{Wang2003}), no definitive answer for which quality measure is best exists since their performance is highly dependent on the given problem. Therefore, within this work we propose the following quality measure (comparative measure only) for a reconstruction result \(a_\text{rec}(x,y)\) of a known defect distribution \(D(x,y)\):
\begin{equation}
    C\left(a_\text{rec}(x,y)\right) = \NMSE\left(D, a_\text{rec}(x,y)\right) + \left\Vert\left(1- \eta^{\prime}(x,y)\right) \odot a_\text{rec}(x,y) \right\Vert_2 \ ,
\end{equation}
where \({\{ \eta^{\prime}(x,y) \in \mathbb{R}: 0\leq\eta^{\prime}(x,y)\leq 1\ \forall x,y\}}\) is the normalized location dependant penalty mask \(\eta(x,y)\) defined as:
\begin{equation}
    \eta(x,y)=D(x,y) *_{x,y} \Phi_\text{PSF}(x,y) \ .
\end{equation}
This measure \(C\left(a_\text{rec}(x,y)\right) \in \left[ 0, \infty \right[\) combines the pixel-wise comparison of the reconstruction with the true defect distribution as performed by the \ac{NMSE} with a location-dependent term \(\eta^{\prime}(x,y)\), which penalizes false positive signals more that are further away from a true defect signal. This factor takes full effect for all false positive signals that are further away from the true position than the spatial width (\(\gtrapprox 3\,\sigma_\text{PSF}\)) of the \ac{PSF}. In this context, smaller values of \(C\left(a_\text{rec}(x,y)\right)\) indicate a better reconstruction result. The \ac{NMSE} is given by the following equation:
\begin{equation}
    \NMSE\left(x_\text{true}, x_\text{rec}\right) = \frac{\left\Vert x_\text{true} - x_\text{rec} \right\Vert_2^2}{\left\Vert x_\text{true}-\overline{x_\text{true}}\right\Vert_2^2} \ .
\end{equation}
With the help of this quality measure it is now possible to define a minimization problem, which maximizes the quality of the reconstruction of the internal heat source distribution \(a_\text{rec}(x,y)\) by choosing the optimal set of regularization parameters \(\Lambda_\text{best}\) as input for \refeqq{eq:min} as follows:
\begin{equation}\label{eq:best_lambda}
    \Lambda_\text{best} = \left(\lambda^\text{best}_{2,1},\, \lambda^\text{best}_2\right) = \argmin_{\lambda_{2,1},\, \lambda_2}\ C\left(a_\text{rec}(x,y)\right)
\end{equation}
This minimization problem stated in \refeqq{eq:best_lambda} is severely ill-posed and computationally very expensive since for every determination of \( C\left(a_\text{rec}(x,y)\right)\) for a suitable candidate for \(\Lambda_\text{best}\) the similarily ill-posed minimization problem stated in \refeqq{eq:min} has to be solved. Therefore, applying a suitable heuristic search algorithm, which is able to efficiently search through the vast solution space is key in finding a (at best optimal) solution in a feasible time frame. While in the past this process has been carried out mostly manually on an empirical basis, in this work we propose the use of the \textit{differential evolution algorithm}~\cite{Storn1997} to find the (optimal) set of regularization parameters \(\Lambda_\text{best}\) in order to automate and speed-up this process significantly. While this search method is very robust and even works with non-differentiable problems since it only heuristically samples the solution space, it is not guaranteed that the optimal solution will be found. While this could possibly lead to insufficient reconstruction quality by settling on a local minimum far from the global one, for a sufficient amount of agents (population size for the heuristic search) \(n_\text{agents}\gg 10\) this has not yet occured to be an issue.

\section*{Experimental Setup}\label{sec:exp_setup}
In order to validate the synthetic measurement data \(T_\text{meas,sim}\), which can be generated by \refeqq{eq:fws} as shown in \reffig{fig:zeta_fit} and in order to assess the overall capabilities of \ac{PSR} reconstruction, several measurements in the lab have been performed. Here, a laser-coupled \ac{DMD} projector based on a DLP650LNIR \ac{DLP} chip from Texas Instruments featuring \({n_\text{pix}=1280\times 800}\) (WXGA, \({16\mathbin{:}10}\)) pixels at a pixel size of \({d_\text{pix, proj}=\SI{10.8}{\micro\metre}}\) has been utilized to project the illumination patterns for each measurement. This projector is coupled to a diode laser, which supplies the maximum necessary optical input power of \({\hat{Q}_\text{optical,in}=\SI{270}{\watt}}\) to the projector at a wavelength of \(\lambda=\SI{940}{\nano\metre}\) resulting in an optical output power of \({\hat{Q}_\text{optical}=\SI{86}{\watt}}\) at \({\beta=1}\). With the attached objective, which features a \(1.85\times\)~magnification, a single pixel size of \({d_\text{pix}=\SI{20}{\micro\metre}}\) and an irradiance on the \ac{OUT} of \(\SI{21}{\watt\per\centi\metre\squared}\) has been achieved. Reaching such high irradiance with a \ac{DLP}-based optical system lies on the upper edge of what is achievable with current \ac{DLP} technology and requires serious cooling efforts within the device in order to savely operate the device. However, for testing metallic materials for defects an irradiance in the order of \(\sim \SI{10}{\watt\per\centi\metre\squared}\) is necessary for sufficient heating.\\

To increase the transfered power per pixel, every 20 pixels have been clustered together within the scope of this work. This leads to a total clustered pixel size of \(d_\text{spix}=\SI{0.4}{\milli\metre}\), which is on the edge of the reasonable pixel size range for the automatic determination of the best regularization parameters (see \reffig{fig:sparsity_patterns}) with the help of the forward solution. While also much smaller pixel cluster sizes have shown good results in the past, deliberately choosing the cluster size this close to the limit for the automatic regularization parameter determination has been performed to further give a hint on the robustness of the method to non-ideal experimental conditions.\\

\begin{figure}[!h]
    \centering
    \ifcomptikz
        \scalebox{1}{\import{figures/}{setup.tex}}
    \else
        \includegraphics[scale=1]{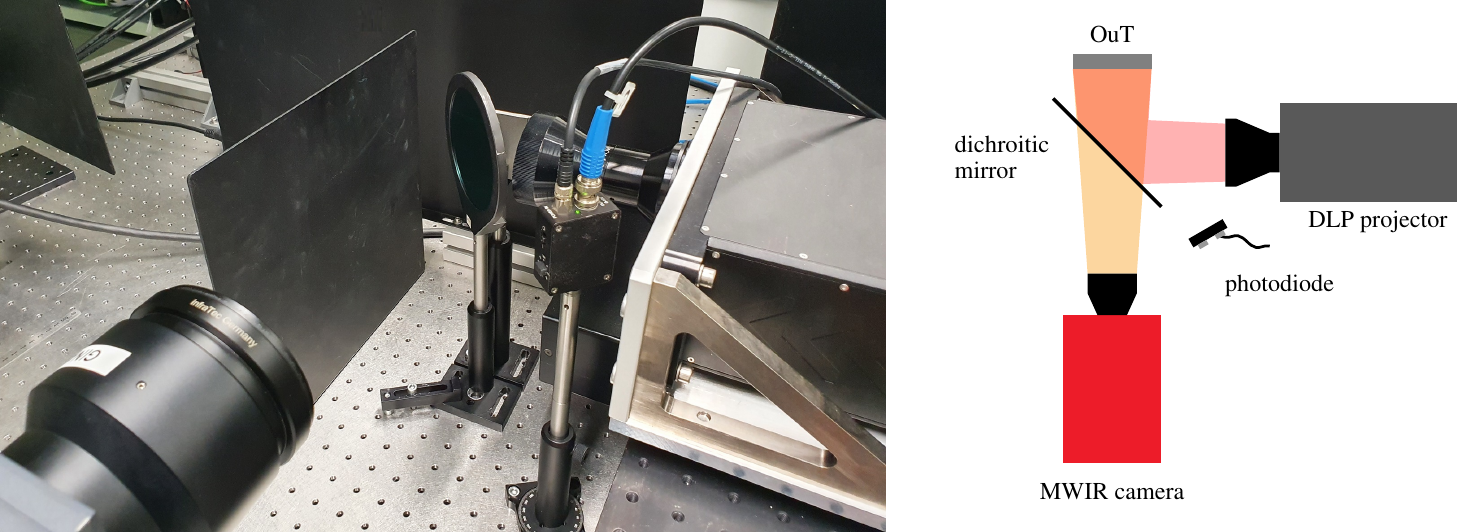}
    \fi
    \caption{Experimental Setup: A laser-coupled \ac{DLP} projector (right) projects different illumination patterns via a dichroic mirror (middle) onto the \ac{OUT} (middle, background) while the resulting change in front surface temperature is recorded via a \ac{MWIR} camera (bottom left). A photodiode (middle, foreground) is detecting when the patterns are projected and triggers the camera to start recording.}\label{fig:exp_setup}
\end{figure}

The resulting experimental setup is shown in \reffig{fig:exp_setup}. Within this particular setup, a dichroic mirror is used to separate the illumination and camera beam paths. This mirror is highly reflective for the laser wavelength while being transparent in the infrared wavelength range. An overview over all experimental parameters is additionally given in \reftab{tab:experimental_params}.\\

The front surface temperature of the \ac{OUT} has been recorded using a cooled \ac{MWIR} camera at a spatial resolution of \({\Delta x, \Delta y = \SI{0.1}{\milli\metre}}\) with a framerate of \({f_\text{cam}=\SI{160}{\hertz}}\) and an NETD of~\({<\SI{50}{\milli\kelvin}}\). The initial temperature \(T_0(x,y)\) has been determined for each individual measurement by averaging 50~frames directly before the illumination has been triggered. The start trigger signal for the camera has been provided by a photodiode sensing the start of the laser pulse used for photothermally heating the \ac{OUT}.\\
\begin{figure}[!h]
    \centering
    \ifcomptikz
        \scalebox{1}{\import{figures/}{sample_ROI.tex}}
    \else
        \includegraphics[scale=1]{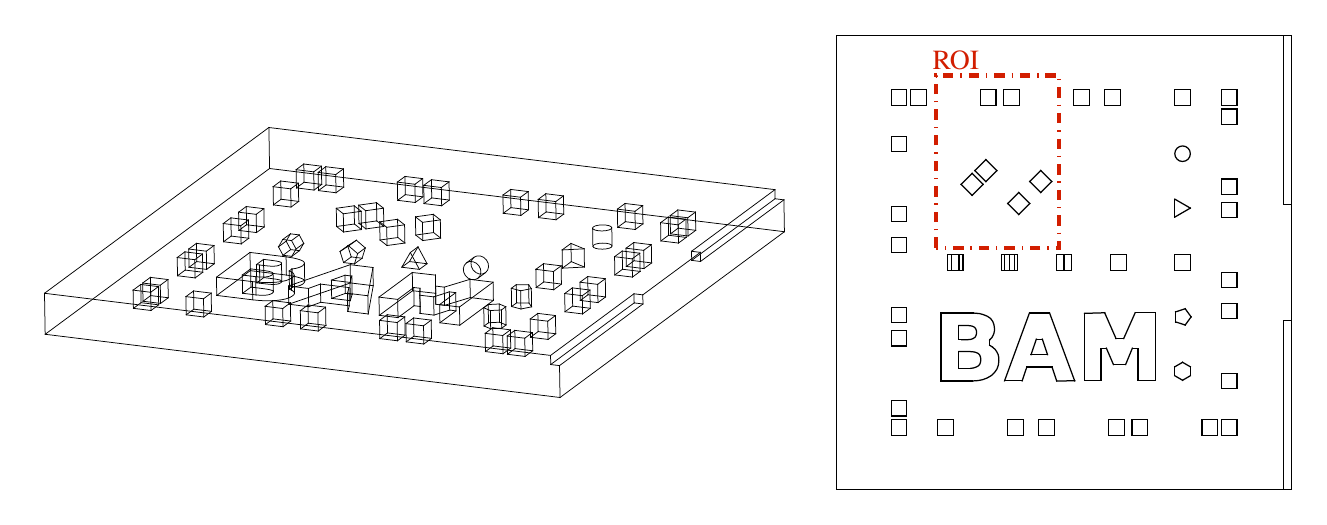}
    \fi
    \caption{\Acl{OUT} and \ac{ROI}: The \ac{OUT} investigated within this work has been additively manufactured from 316L stainless steel featuring several cubical internal defects with side lengths \(d_\text{defect}=\SI{2}{\milli\metre}\) of unfused metal powder lying \(\SI{0.5}{\milli\metre}\) below the front surface. The \ac{ROI} considered in this work encompasses three defect pairs with spacings \(\{0.5, 1, 2\}\si{\milli\metre}\) with two defect pairs oriented at \(\SI{45}{\degree}\) to the illumination pattern grid and the sensor pixel grid of the \ac{MWIR} camera. Left: isometric wire frame view of the \ac{OUT}, Right: front view (wire frame) with \ac{ROI}.}\label{fig:sample_ROI}
\end{figure}

The square platelike \ac{OUT} examined within this work and the corresponding \ac{ROI} is shown in \reffig{fig:sample_ROI}. It features a side length of \(\SI{58.5}{\milli\metre}\) and a thickness of \(L=\SI{4.5}{\milli\metre}\) and has been additively manufactured from 316L stainless steel (\({k=\SI{15}{\watt\per\metre\per\kelvin}}\), \({\rho=\SI{7950}{\kilo\gram\per\metre\cubed}}\), \({c_p=\SI{502}{\joule\per\kilo\gram\per\kelvin}}\), \({\alpha=\SI{3.76e-6}{\metre\squared\per\second}}\))~\cite{International2009,EGS2019} and features cubical internal defects with side length \(d_\text{defect}=\SI{2}{\milli\metre}\) starting at a depth of \(\SI{0.5}{\milli\metre}\) filled with residual unfused metal powder from the manufacturing process. The chosen \ac{ROI} spans an area of \(\SI{24.8}{\milli\metre}\, \times\, \SI{15.5}{\milli\metre}\) and encompasses three defect pairs with separation distances of \(0.5\), \(1\) and \(\SI{2}{\milli\metre}\). This variance in defect spacing allows to assess the resolution capabilities of the \ac{PSR} reconstruction by means of determining for which separation distance can individual defects still be identified as separate defects. Furthermore, two of the three defect pairs are oriented at \(\SI{45}{\degree}\) with respect to the illumination pattern pixel grid and the pixel grid of the \ac{MWIR} camera. This further benchmarks the capabilities of the algorithm and gives a strong hint about the independence of the algorithm from defect orientation.\\

In order to not introduce any history in the measurements, all individual illuminations have been performed with a laser pulse length of \({t_\text{pulse}=\SI{0.5}{\second}}\) and a conservative delay of \(\SI{25}{\second}\) between subsequent illuminations allowing the \ac{OUT} to cool back down to ambient temperature. Therefore, the measurement of a \ac{ROI} with \({n_m=20}\) patterns will be completed in about \(\SI{8.3}{\minute}\). In comparison, this represents a drastic decrease in measurement time compared to sequential spot-wise illumination based \ac{PSR} where measuring an equally sized \ac{ROI} with similar delay between measurements would take approximately \(\SI{2}{\hour}\) to complete. Optimizing the delay between measurements or even remove any history from the measurements by subtracting the total increase in temperature over time obtained from a suitable modeling of the \acg{OUT} temperature evolution would still be possible to further speed-up the measurement process.

\begin{longtable}[c]{P{2.5cm} l P{1cm}r @{\:} l}
    \caption{Overview over the experimental parameters.}\label{tab:experimental_params}                                                                                                                              \\
                                                                & \textbf{Parameter}                  &                            & \multicolumn{2}{c}{\textbf{Value}}                                              \\\midrule\endfirsthead
    \caption[]{Overview over the experimental parameters: continued.}                                                                                                                                                \\
                                                                & \textbf{Parameter}                  &                            & \multicolumn{2}{c}{\textbf{Value}}                                              \\\midrule\endhead
    \multirow{4}{=}{\centering\textbf{\ac{OUT}}}                & \ac{ROI}                            &                            & \(\num{24.8}\, \times\, \num{15.5}\) & \(\si{\milli\metre\squared}\)            \\*
                                                                & thickness                           &                            & \(\num{4.5}\)                        & \(\si{\milli\metre}\)                    \\*
                                                                & material                            &                            & 316L stainless                       & steel                                    \\*
                                                                & thermal diffusivity                 & \(\alpha\)                 & \(\num{3.76}\)                       & \(\si{\milli\metre\squared\per\second}\) \\[0.4em]
    \multirow{3}{=}{\centering\textbf{defect pattern}}          & defect side length                  &                            & \(\num{2}\)                          & \(\si{\milli\metre}\)                    \\*
                                                                & starting depth                      &                            & \(\num{0.5}\)                        & \(\si{\milli\metre}\)                    \\*
                                                                & spacings                            &                            & \(\{0.5, 1, 2\}\)                    & \(\si{\milli\metre}\)                    \\[0.4em]
    \multirow{3}{=}{\centering\textbf{infrared camera}}         & spatial resolution                  & \(\Delta x, \Delta y\)     & \(\num{0.1}\)                        & \(\si{\milli\metre}\)                    \\*
                                                                & acquisition frequency               & \(f_\text{cam}\)           & \(\num{160}\)                        & \(\si{\hertz}\)                          \\*
                                                                & NETD                                &                            & \(<\num{50}\)                        & \(\si{\milli\kelvin}\)                   \\[0.4em]
    \multirow{5}{=}{\centering\textbf{illumination parameters}} & optical output power at \(\beta=1\) & \(\hat{Q}_\text{optical}\) & \(\num{86}\)                         & \(\si{\watt}\)                           \\*
                                                                & irradiance at \ac{ROI}              &                            & \(\num{21}\)                         & \(\si{\watt\per\centi\metre\squared}\)   \\*
                                                                & pixel size at \ac{ROI}              & \(d_\text{pix}\)           & \(\num{20}\)                         & \(\si{\micro\metre}\)                    \\*
                                                                & pixel cluster size                  & \(d_\text{spix}\)          & \(\num{0.4}\)                        & \(\si{\milli\metre}\)                    \\*
                                                                & pulse length                        & \(t_\text{pulse}\)         & \(\num{0.5}\)                        & \(\si{\second}\)                         \\[0.4em]
\end{longtable}
\FloatBarrier
\section*{Results}\label{sec:exp_res}
\begin{figure}[!h]
    \centering
    \ifcomptikz
        \scalebox{1}{\import{figures/}{result_rec.tex}}
    \else
        \includegraphics[scale=1]{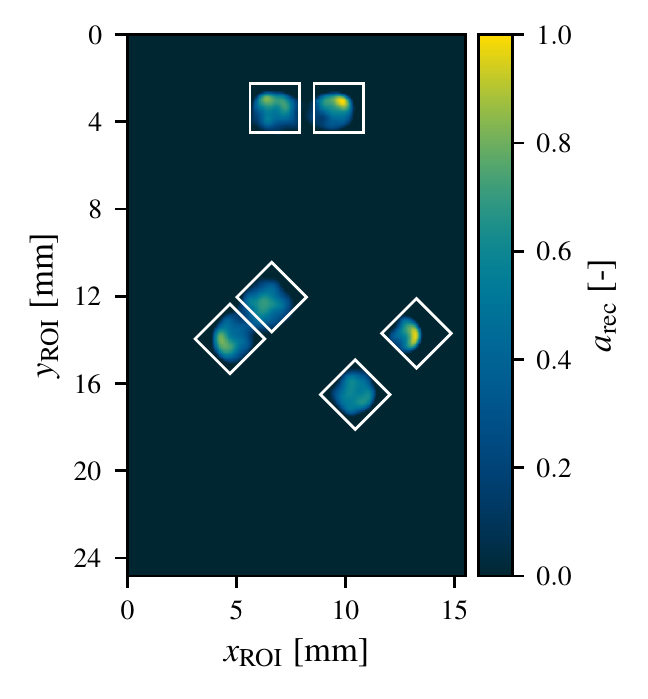}
    \fi
    \vspace{-1em}
    \caption{Reconstruction result \(a_\text{rec}(x,y)\) obtained for \({\Lambda_{best}=\left(490, 34.4\right)}\) with \({\rho_\text{ADMM}=9900}\) for an illumination with \({n_m=20}\) patterns with \({d_\text{spix}=\SI{0.4}{\milli\metre}}\) and \({\beta=0.5}\) after \({n_\text{iter}=100}\) iterations. The true defect postions are indicated by white boxes. All defects have been detected with a reasonable reconstruction of the defect shape.}\label{fig:result_rec}
\end{figure}

For projecting \({n_m=20}\)~patterns with \({\beta=0.5}\) and subsequent \ac{PSR} reconstruction using the automatic regularization parameter determination, the reconstruction result as shown in \reffig{fig:result_rec} has been achieved. For the determination of \({\Lambda_{best}=\left(490, 34.4\right)}\) within the scope of the differential evolution algorithm applied, 549~reconstruction problems as stated in \refeqq{eq:min} have been solved iteratively (\({n_\text{iter}=100}\)~each) without any additional user input, which on modern computer hardware took about \(\SI{1.5}{\hour}\) to perform.

As can be seen in the reconstruction result, all defects have been detected with no false positive defect signals showing up. Furthermore, all defects are nicely detectable as separate, while only the defect pair with the smallest spacing shows room for improvement in this regard. Using a smaller pixel cluster size and increasing the number of patterns projected are expected to further improve the reconstruction for smaller spacings~\cite{Lecompagnon2022a}. The overall reconstruction quality of the defect shape has been also quite decent only with the right-most defect not fully reconstructed near the edge of the \ac{ROI}. This can be most likely explained by it being very close to the edge of the \ac{ROI} and therefore there is missing information for this part of the defect compared to all others. Even though all defect pairs consist of identical cubical defects, the reconstruction of each of them shows some variation when compared between each other. While this is not ideal, it can be also traced back to the fact that the necessary condition for a homogeneous reconstruction~(cf.~\refeqq{eq:SRcond}) is more and more violated close to the edge of the \ac{ROI} and is also expected to improve with the number of measurements performed.\\

\begin{figure}[!h]
    \centering
    \ifcomptikz
        \scalebox{1}{\import{figures/}{meth_comp.tex}}
    \else
        \includegraphics[scale=1]{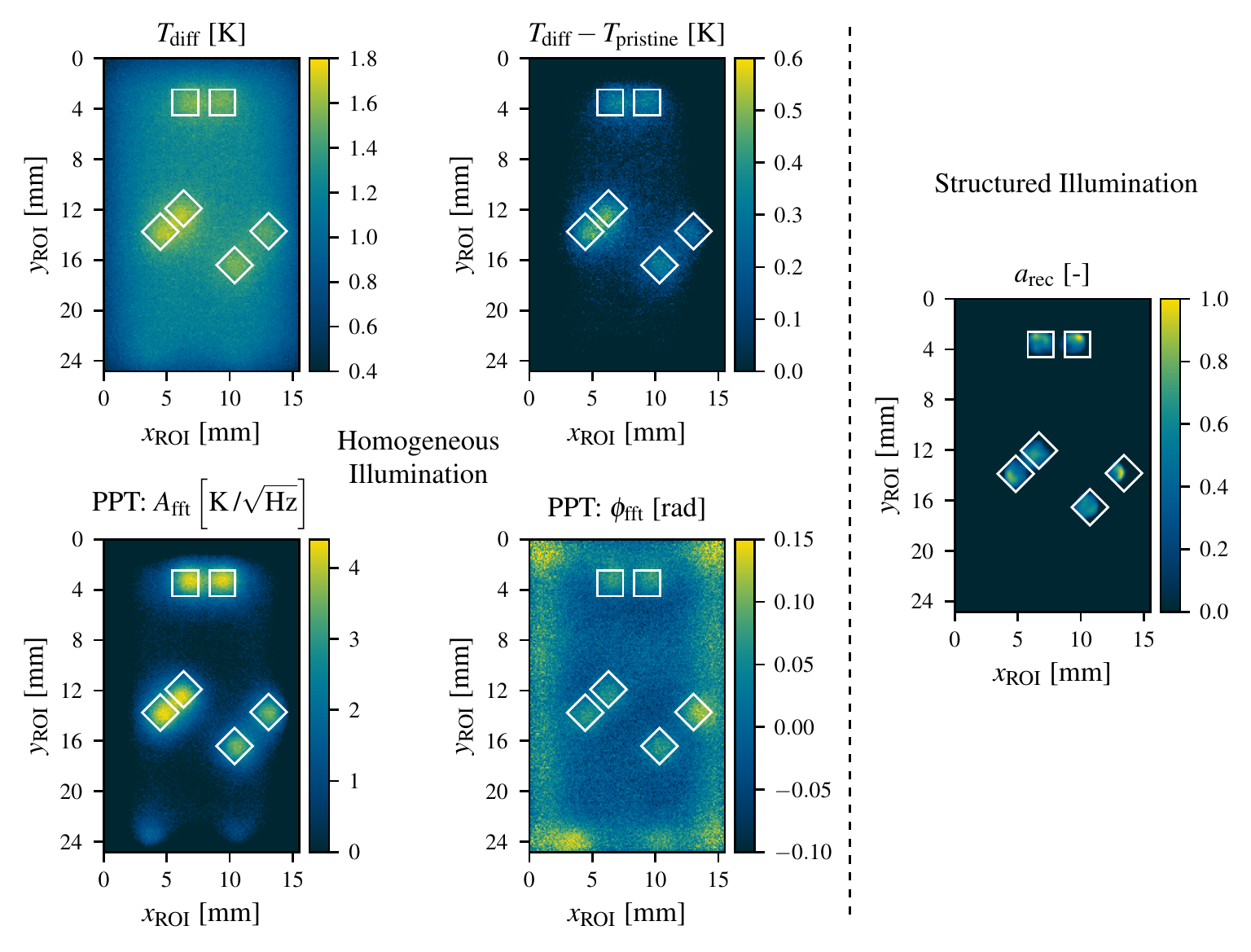}
    \fi
    \vspace{-1em}
    \caption{Qualitative comparison of different defect detection methods: The left four plots show results obtained using conventional detection methods with homogeneous illumination. The top left plot contains the temperature increase \(T_\text{diff}\) obtained at \(t_\text{eval}=t_\text{pulse}=\SI{0.5}{\second}\). The top right plot shows a difference thermogram where from each pixel the expected increase in temperature for a pristine subarea of the \ac{ROI} is subtracted. The bottom two plots display the amplitude and phase image for a \ac{PPT} evaluation on \(T_\text{diff}\) for a frequency of \(f_\text{PPT}=\SI{0.516}{\hertz}\). On the right, the \ac{PSR} reconstruction result as obtained in \reffig{fig:result_rec} is depicted. The true defect positions are indicated by white boxes.}\label{fig:result_comp}
\end{figure}

\begin{figure}[!h]
    \centering
    \ifcomptikz
        \scalebox{1}{\import{figures/}{method_comp_cut.tex}}
    \else
        \includegraphics[scale=1]{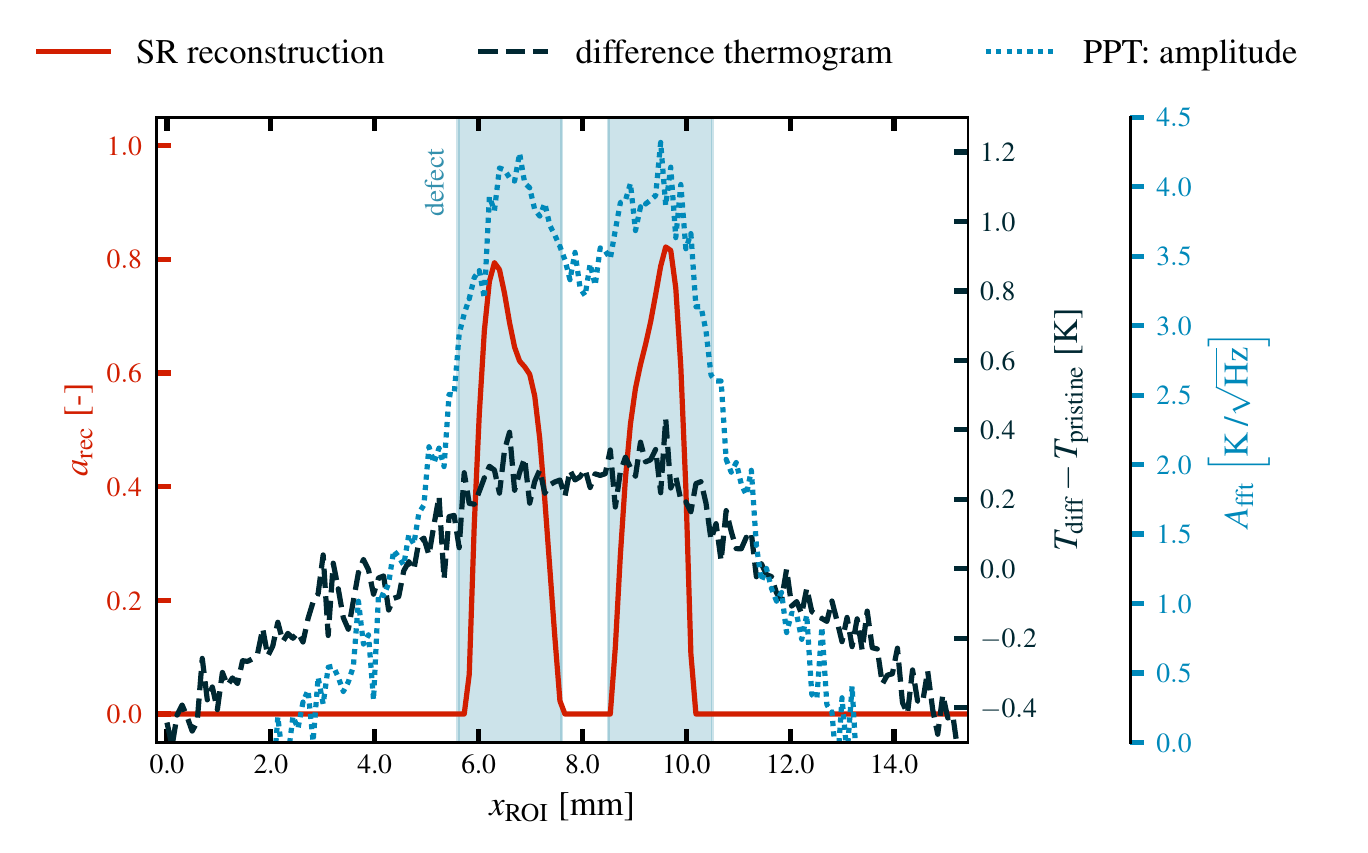}
    \fi
    \vspace{-1em}
    \caption{Sectional view of the comparison of the results of different defect detection methods over a defect pair at \({y_\text{ROI}=\SI{2.81}{\milli\metre}}\). Here, the (sparse) \ac{PSR} reconstruction result as obtained in \reffig{fig:result_rec} is shown in conjunction with the results from the difference thermogram and the amplitude of the \ac{PPT} evaluation. }\label{fig:result_comp_cut}
\end{figure}

In order to evaluate the quality of the reconstruction as shown in \reffig{fig:result_rec}, a comparison to the results of several other established thermographic defect detection techniques using a single homogeneous illumination of the \ac{ROI} has been performed. Due to the lack of a suitable universally applicable measure for reconstruction quality, this comparison has only been carried out in a qualitative fashion and the results are shown in \reffig{fig:result_comp}. Since \ac{PSR} reconstruction results in a sparse defect map compared to (most) other methods, which give out continuous data, a quantitative comparison of the defect reconstruction quality is a highly non-trivial task. This fact is further emphasized in \reffig{fig:result_comp_cut} in which a sectional view of the results of the different methods is presented.

In this comparison the reconstruction result as shown in \reffig{fig:result_rec} is displayed side-by-side with conventional methods. These methods consist of the difference thermogram \(T_\text{diff}\) for \(t_\text{eval}=t_\text{pulse}=\SI{0.5}{\second}\), which features the maximum defect contrast. In addition, \(T_\text{diff}\) for which additionally the expected temperature for a defect free region of the \ac{ROI} has been subtracted is displayed. Furthermore, \ac{PPT} has been performed on the whole image sequence and the phase \(\phi_\text{fft}\) and amplitude \(A_\text{fft}\) images have been evaluated for a frequency of \(f_\text{PPT}=\SI{0.516}{\hertz}\,\)~\cite{IbarraCastanedo2004}. For a fair comparison, the homogeneous illumination of the \ac{ROI} has been performed in the same setup as the measurements resulting in the reconstruction from \reffig{fig:result_rec} using a fully activated image (\(\beta=1\)) at maximum output power of the projector of \({\hat{Q}_\text{optical}=\SI{86}{\watt}}\).

\section*{Conclusion}\label{sec:end}
In the comparisons shown in \reffig{fig:result_comp} and \reffig{fig:result_comp_cut} the added benefit of the \ac{PSR} reconstruction technique can be clearly observed. Not only are all defects clearly visible for a human, the nature of the signal as a sparse defect pattern differentiating clearly between defect-free and defective regions allows for a clear labeling of the \ac{ROI}, which can be interpreted by humans even with little training in the subject as well as it is easily usable for further automatic data processing by machines. This advantage of the proposed experimental \ac{PSR} reconstruction approach can be clearly observed when comparing the obtained results with other sophisticated thermographic internal defect resolution methods making use of spatially structured illumination like photothermal coherence tomography~\cite{Kaiplavil2014,Tavakolian2017} or thermal wave slice diffraction tomography~\cite{Nicolaides2000}. In addition, the application of fully two-dimensionally structured random pixel patterns has shown to lead to a drastic decrease in measurement times compared to the current state of the art sequential point-wise illumination strategies applied for \ac{PSR} reconstruction~\cite{Ahmadi2021,Lecompagnon2022}. Nevertheless, the increased experimental complexity and measurement times compared to the also shown conventional methods still poses a challenge for the application in large volume production but for high-reliability applications in medicine or for the testing of aerospace products, the added quality and resolution capabilities of the method clearly outweigh the increased measurement times. The absence of an automatic determination method of a suitable set of regularization parameters is currently still the largest drawback when working with this testing method. While within this work we have shown a way to achieve this automatic determination with prior knowledge of the defect structure, this inversion method is still not very suitable for real word testing scenarios and more aimed towards scientific research on the subject. However, there is current ongoing work in order to solve this issue using machine learning techniques~\cite{Hauffen2022,Ahmadi2022}. The introduction of a forward solution within this work also helps out to better tune the reconstruction parameters since it is now technically feasible to generate varying synthetic datasets for further exploration of the capabilities of the method. Furthermore, the maximum optical output power of the \ac{DLP}-projector applied in the experimental section of this work has been observed to be still lacking to detect defects in materials with high thermal conductivity (stainless steel in this case) than what has been presented in this work even though the \ac{DMD}-chip in this projector is currently a top-of-the-line model with regards to achievable output power. Here, a further improvement in \ac{DLP}-projector technology is still necessary.
\FloatBarrier
\section*{Data Availability}
The data that support the findings of this study are available from the corresponding author upon reasonable request.

\section*{Declaration of interests}
The authors declare that they have no known competing financial interests or personal relationships that could have appeared to influence the work reported in this paper.

\section*{Author contributions statement}
J.L. and M.Z. conceptualized the idea for 2D \acl{PSR}, J.L. designed the specimen, developed the methodology and all special software used. P.D.H. and J.L. carried out the experiments in the lab, J.L. wrote the manuscript with support from M.Z., C.R. and M.Z. aided with crucial scientific supervision. All authors reviewed the manuscript.

\bibliography{refs}

\end{document}